\documentclass[conference]{IEEEtran}
\IEEEoverridecommandlockouts
\usepackage{cite}
\usepackage{amsmath,amssymb,amsfonts}
\usepackage{algorithmic}
\usepackage{graphicx}
\usepackage{textcomp}
\usepackage{xcolor}

\usepackage{algorithm}
\usepackage{cases} 
\usepackage{enumerate}
\usepackage{bm}

\begin{document}

\title{A Multi-Agent Deep Reinforcement Learning based Spectrum Allocation Framework for D2D Communications\\
\thanks{This work was supported by the National Natural Science Foundation of China under Grant No. 61571062 and No. 61871047.}
}

\author{\IEEEauthorblockN{Zheng Li\IEEEauthorrefmark{1},
		Caili Guo\IEEEauthorrefmark{2}, Yidi Xuan\IEEEauthorrefmark{1}}
	\IEEEauthorblockA{\IEEEauthorrefmark{1}Beijing Laboratory of Advanced Information Networks,\\Beijing University of Posts and Telecommunications, Beijing, China 100876\\ }
	\IEEEauthorblockA{\IEEEauthorrefmark{2}Beijing Key Laboratory of Network System Architecture and Convergence, \\Beijing University of Posts and Telecommunications, Beijing, China 100876\\ 
		Email: lizhengzachary@bupt.edu.cn}}

\maketitle

\begin{abstract}
Device-to-device (D2D) communication has been recognized as a promising technique to improve spectrum efficiency. However, D2D transmission as an underlay causes severe interference, which imposes a technical challenge to spectrum allocation. Existing centralized schemes require global information, which can cause serious signaling overhead. While existing distributed solution requires frequent information exchange between users and cannot achieve global optimization. In this paper, a distributed spectrum allocation framework based on multi-agent deep reinforcement learning is proposed, named Neighbor-Agent Actor Critic (NAAC). NAAC uses neighbor users' historical information for centralized training but is executed distributedly without that information, which not only has no signal interaction during execution, but also utilizes cooperation between users to further optimize system performance. The simulation results show that the proposed framework can effectively reduce the outage probability of cellular links, improve the sum rate of D2D links and have good convergence.
\end{abstract}


\section{Introduction}
D2D communication allows two nearby users to form a D2D pair and communicate with each other directly, thus improving the transmit quality significantly due to short transmission distance \cite{kai2019resource}. D2D underlay communication reuses the spectrum of the cellular network to potentially increase spectral efficiency. However, D2D communication generates interference to the cellular network if the radio resources are not properly allocated \cite{feng2013device}. Thus, it is necessary to study the resource allocation to ensure the reliability of cellular communication and increase the network capacity.

There are many resource allocation methods in the existing literature, which can be divided into centralized and distributed schemes. In the centralized schemes\cite{wu2018high,kose2018resource,kuang2019energy}, the BS is responsible for allocating resources to the CUEs and D2D pairs, and monitoring information such as Signal to Interference-plus-Noise Ratio (SINR), channel state information (CSI), etc. However, as the number of users increases, acquiring global CSI will cause severe signaling overhead, and a centralized scheme for spectrum allocation becomes unrealistic. Moreover, the complexity of centralized algorithms increases with the number of users, causing enormous computational pressure on the BS.

In order to reduce the signaling overhead and alleviate the computational pressure of the BS, a series of distributed methods are proposed. In a distributed approach, without a central controller, the D2D pairs opportunistically reuse the spectrum of cellular users (CUEs). Distributed schemes can scale well to larger networks, but require frequent exchange of information between adjacent D2D users. Quite a few distributed algorithms are based on game theory. \cite{zaki2017distributed} models D2D pairs sharing spectrumc with CUEs as an auction mechanism. Large signaling overhead is incurred since the CSI and prices have to be shared among the D2D pairs. Moreover, this type of method requires a lot of iteration to converge.

In addition to game theory, machine learning has been considered as a effective tool in solving different network problems in 5G \cite{jiang2017machine}. Reinforcement learning (RL) is one of the most powerful machine learning tools for policy control \cite{sutton1998introduction}. Recently, many works have applied reinforcement learning to solve the intelligent resource management problem in D2D underlay networks \cite{zia2019distributed,yang2019intelligent,ye2019deep}. Each D2D pair is supported by an agent, which automatically selects a reasonable spectrum based on the policy learned by reinforcement learning. A Q-learning based resource allocation is proposed in \cite{zia2019distributed}. Since Q-learning isn't suitable to deal with continuous valued state and action spaces, an actor-critic (AC) approach is proposed in \cite{yang2019intelligent}. In \cite{ye2019deep}, a decentralized mechanism based on deep reinforcement learning has been developed. The above previous works model the policy search process in reinforcement learning as a Markov decision process (MDP). However, in the decentralized settings of spectrum allocation problem, all agents (D2D pairs) are independently updating their policies, which is a multi-agent environment. The environment appears non-stationary from the view of any one agent. 

In this paper, we propose a distributed spectrum allocation framework based on multi-agent deep reinforcement learning, named Neighbor-Agent Actor Critic (NAAC). NAAC is a framework of centralized training with decentralized execution, which uses information from neighbor users for training to effectively overcome the instability of the multi-agent environment, while makes full use of cooperation between users to further improve system performance. NAAC does not require information interaction when it is executed, so it significantly saves signaling overhead. The main contributions of this paper are summarized as follows:
1. The multi-agent environment is modeled by Markov game, which is more accurate and helpful for the study of subsequent reinforcement learning algorithms than MDP.
2. Proposing a multi-agent reinforcement learning framework of centralized training with decentralized execution, which solves the problem that the multi-agent environment is unstable and leverages the partnership between users to further increase system performance.

The rest of this paper can be organized as follows. Section
II shows the system model and formulates the optimization problem. In Section III, we model the resource management problem as a partially observable Markov games and the NAAC framework is adopted to address it. The simulation results and analysis are presented in Section VI. Finally, Section V concludes the paper.

\section{System Model and Problem Formulation}

\subsection{System Model}

An downlink scenario in a single cell system is considered. A set of $ M $ CUEs, denoted as $ \mathcal{M} = \{1,...,M\} $ and a set of $ N $ active D2D pairs, denoted as $ \mathcal{N} = \{1,...,N\} $ are located in the coverage area of one BS. We denote the $ m^{th} $ CUE in the system by $ C_{m} $, $ m \in \mathcal{M} $, the $ n^{th} $ D2D pair by $ D_{n} $, $ n \in \mathcal{N} $, the transmitter and the receiver of D2D pair $ D_{n} $ by $ D^{t}_{n} $ and $ D^{r}_{n} $, respectively. Orthogonal Frequency Division Multiple Access (OFDMA) is employed to support multiple access for both the cellular and D2D communications, where a set of $ K $ resource blocks (RBs) are available for spectrum allocation. In this system, the D2D pairs share the same spectrum with the CUEs.

We assume that the BS and the transmitter of a D2D pair transmit with power $ P^{b} $ and $ P^{d} $, respectively. The channel gains of the cellular communication link from the BS to CUE $ C_{m} $, the D2D communication link from D2D transmitter $ D^{t}_{n} $ to D2D receiver $ D^{r}_{n} $, the interference link from D2D transmitter $ D^{t}_{n} $ to CUE $ C_{m} $, the interference link from the BS to D2D receiver $ D^{r}_{n} $ and the interference link from D2D transmitter $ D^{t}_{i} $ to D2D receiver $ D^{r}_{n} $ when they share the same spectrum for data transmission, are represented by $ g^{b,c}_{m} $, $ g^{t,r}_{n} $, $ g^{t,c}_{n,m} $, $ g^{b,r}_{n} $ and $ g^{t,r}_{i,n} $, respectively.  The energy of the additive white Gaussian noise (AWGN) at a receiver is denoted by $ \sigma^{2} $.

The instantaneous SINR of the received signal at CUE $ C_{m} $ from the BS in RB $ k $ can be written as
\begin{equation}
\xi^{c}_{m,k}
=\dfrac{P^{b}g^{b,c}_{m}}
{\sum\limits_{n \in \mathbf{D}_{k}}P^{d}g^{t,c}_{n,m}
	+\sigma^{2}}, 
\label{cue_sinr}
\end{equation}	
where $ \mathbf{D}_{k} $ represents the set of D2D pairs to which RB $ k $ is allocated. The instantaneous SINR of the received signal at the $ D^{r}_{n} $ from $ D^{t}_{n} $ in RB $ k $ can be written as
\begin{equation}
\xi^{d}_{n,k}
=\dfrac{P^{d}g^{t,r}_{n}}
{P^{b}g^{b,r}_{n}
	+\sum\limits_{i \in \mathbf{D}_{k},i \ne n}
	P^{d}g^{t,r}_{i,n}
	+\sigma^{2}}.
\label{d2d_sinr}
\end{equation}

\subsection{Problem Formulation} \label{Problem Formulation}
We assume that each CUE has been assigned a RB and a RB can be allocated to multiple D2D pairs. We define a RB allocation matrix $ \mathbf{A}_{N \times K} = [a_{n,k}] $ for the D2D pairs. When RB $ k $ is allocated to $ D_{n} $, $ a_{n,k} = 1 $, otherwise $ a_{n,k} = 0 $.

Our objective is to find a RB allocation matrix $ \mathbf{A}_{N \times K} $ for all D2D pairs to maximize the D2D sum rate, which can be formulated as

\begin{equation}
\max_{\mathbf{A}_{N \times K} = [a_{n,k}]} 
\sum\limits_{n=1}^N \sum\limits_{k=1}^K a_{n,k} \log _{2}(1+\xi^{d}_{n,k}),
\label{optimal_problem}
\end{equation}
\begin{numcases}{s.t.  }
\sum\limits_{k=1}^K a_{n,k} \leq 1, a_{n,k} \in \{0,1\}, \forall n \in \mathcal{N}, k \in \mathcal{K}
\label{allocation_constraint} \\
\xi^{c}_{m,k} \geq \xi^{c}_{min}, \forall m \in \mathcal{M}, k \in \mathcal{K}
\label{cue_sinr_constraint}
\end{numcases}
Constraints in (\ref{allocation_constraint}) imply that a maximum of one RB can be allocated to each D2D pair. Since CUEs are the primary users of the frequency band, the transmission quality of CUEs should be ensured, constraints in (\ref{cue_sinr_constraint}) imply that the SINR of CUEs should satisfy a predefined constraint, where $ \xi^{c}_{min} $ represents the SINR threshold of the CUE.

\section{Neighbor-Agent Deep Reinforcement Learning based Spectrum Allocation} \label{NADreinforcement learning}
The optimization problem in (\ref{optimal_problem}) is difficult to solve as it is a NP-hard combinatorial optimization problem \cite{plaisted1976some}. In addition, we assume that each D2D pair can only obtain its own CSI, and there is no information interaction between D2D users. Using a distributed method to solve this optimization problem requires each D2D pair to choose RB autonomously. The reinforcement learning method is an effective method to solve such problems. Hence, in this section, we first model the multi-agent environment and then a distributed framework based on multi-agent reinforcement learning is proposed to address the spectrum allocation problem.

\begin{figure}[!t]
	\centering
	\includegraphics[width=3.5in]{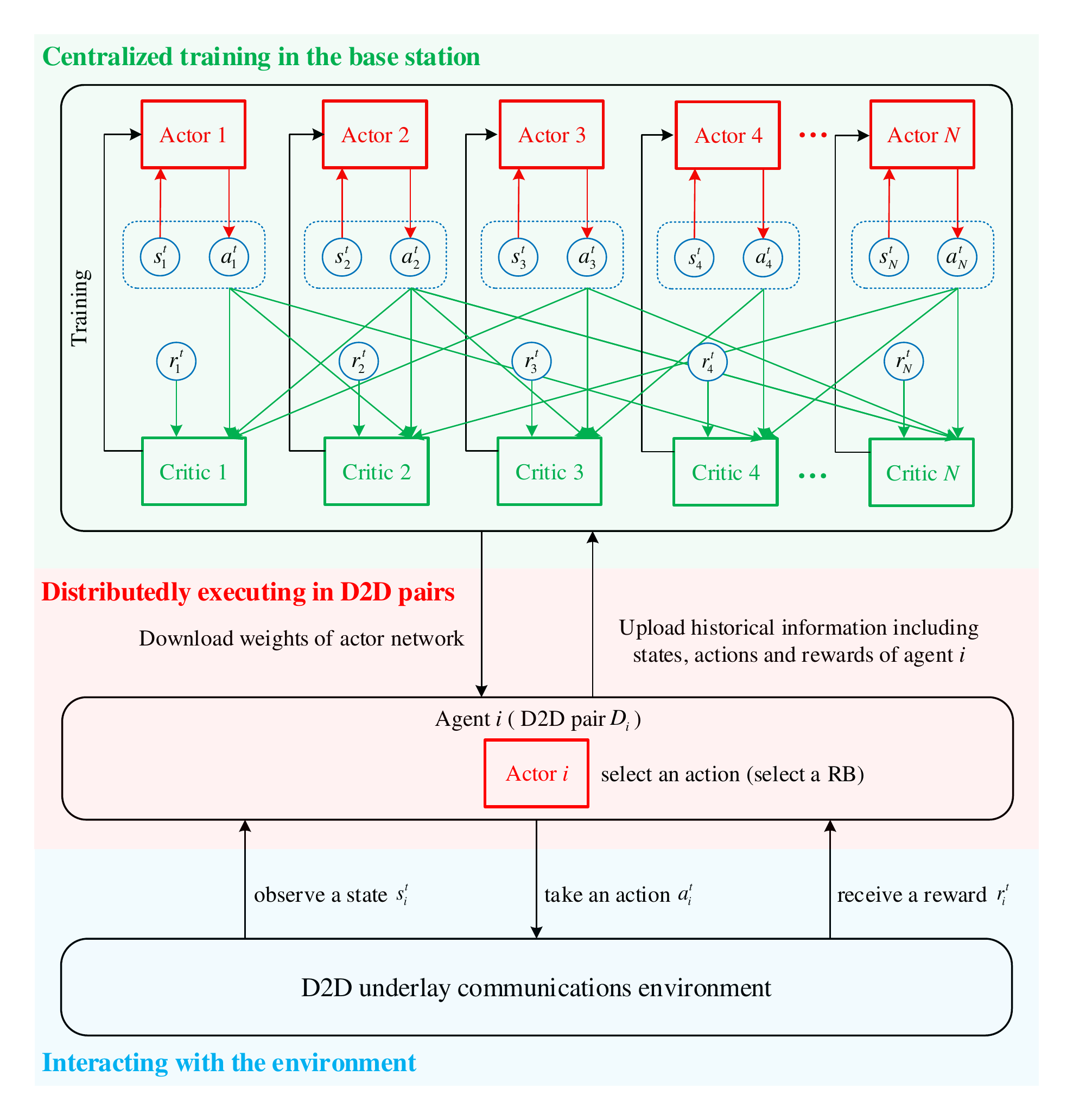}
	\caption{NAAC framework for spectrum allocation in D2D underlay communications.}
	\label{NAAC}
\end{figure}

\subsection{Modeling of Multi-Agent Environments}

Reinforcement learning is an area of machine learning, which is about taking suitable action to maximize reward in a particular situation. In reinforcement learning for spectrum allocation in D2D underlay communications, an agent corresponding to a D2D pair, interacts with the environment. In this scenario, the environment is considered to be everything outside the D2D link. At each time $ t $, the D2D pair, as the agent, observes a state, $ s^{t} $, from the state space, $ \mathcal{S} $, and accordingly takes an action, $ a^{t} $, from the action space, $ \mathcal{A} $, selecting RB based on the policy, $ \pi $. Following the action, the state of the environment transits to a new state $ s^{t+1} $ and the agent receives a reward, $ r^{t} $.

Most of the existing works model the policy search process in reinforcement learning as a MDP. A MDP can be defined as a tuple $ (\mathcal{S},\mathcal{A},r^{t},p,\gamma) $, $ p $ is the transition probability $ p(s^{t+1}|s^{t}, a^{t}) $ when the agent takes the action $ a^{t} \in \mathcal{A} $ from the current state $ s^{t} \in \mathcal{S} $ to a new state $ s^{t+1} \in \mathcal{S} $, $ \gamma \in [0, 1) $ is a discount factor. The return from a state is defined as the sum of discounted future reward $ R^{t} = \sum_{n=0}^{T} \gamma^{n} r^{t+n} $, where $ T $ is the time horizon. However, in the decentralized settings of spectrum allocation problem, all D2D pairs as agents are independently updating their policies as learning progresses, which is a multi-agent environment, the environment appears non-stationary from the view of any one agent, violating Markov assumptions required for convergence of reinforcement learning.

To make up for the shortcomings of MDP, in this work, we consider a multi-agent extension of MDP called partially observable Markov games to model the multi-agent reinforcement learning. At each time $ t $, the D2D pair $ D_{i} $, as the agent $ i $, observes a state, $ s^{t}_{i} $ , from the state space, $ \mathcal{S} $, and accordingly takes an action, $ a^{t}_{i} $, from the action space, $ \mathcal{A} $, selecting RB based on the policy $ \pi_{i} $.  Following the action, the state of the environment observed by agent $ i $ transits to a new state $ s_{i}^{t+1} $ and the agent receives a reward, $ r_{i}^{t} $.

An N-agent Markov game is formalized by a tuple $ (\mathcal{S},\mathcal{A},r_{1}^{t},...,r_{N}^{t},p,\gamma) $, $ p $ is the transition probability $ p(s_{i}^{t+1}|s_{i}^{t}, a^{t}_{1},...,a^{t}_{N}) $ when all agents take actions $ \{a_{i}^{t} \in \mathcal{A}, i \in \mathcal{N} \} $ from the current state $ s_{i}^{t} \in \mathcal{S} $ to a new state $ s_{i}^{t+1} \in \mathcal{S} $, the constant $ \gamma \in [0, 1) $ represents the reward discount factor across time. The return of agent $ i $ from a state is defined as the sum of discounted future reward
$ R^{t}_{i} = \sum_{n=0}^{T} \gamma^{n} r^{t+n}_{i} $, where $ T $ is the time horizon.

In our system, the state space $ \mathcal{S} $, the action space $ \mathcal{A} $, and the reward function $ r_{i}^{t} $ is defined as follows:

\textbf{State space:} The state observed by the D2D link $ D_{i} $ (agent $ i $) for characterizing the environment consists of several parts: the instant channel information of the D2D corresponding link, $ G_{i}^{d,t} $, the channel information of the cellular link, e.g., from the BS to the D2D transmitter, $ G_{i}^{c,t} $, the previous interference to the link, $ I_{i}^{t-1} $, the RB selected by the D2D link in the previous time slot, $ K_{i}^{t-1} $. Hence the state can be expressed as $ s_{i}^{t} = [G_{i}^{d,t}, G_{i}^{c,t}, I_{i}^{t-1}, K_{i}^{t-1}] $.

\textbf{Action space:} At each time $ t $, the agent $ i $ takes an action $ a^{t}_{i} \in \mathcal{A} $, which represents the agent select a RB, according to the current state, $ s^{t}_{i} \in \mathcal{S} $, based on the decision policy $ \pi_{i} $. The dimension of the action space is $ K $ if there are $ K $ RBs.

\textbf{Reward function:} The learning process is driven by the reward function in the reinforcement learning. Each agent makes decision by maximizing its reward with the interactions of the environment. In order to maximize the D2D sum rate while guaranteeing the transmission quality of CUEs, we design a reward function relating to two parts: the D2D link rate and the SINR constraints of CUE. In our settings, the reward remains positive if the SINR constraints are satisfied; if the constraints are violated, it will be a negative reward, $ r_{neg} < 0 $. The positive reward is proportion to the D2D link rate $ \mathcal{R}_{i}^{d,t} $. We use the Shannon capacity model to evaluate $ \mathcal{R}_{i}^{d,t} = \log (1 + \xi^{d,t}_{i}) $, where $ \xi^{d,t}_{i} $ is the instantaneous SINR of the received signal at the D2D receiver $ D^{r}_{i} $ at time slot $ t $. 

\subsection{Neighbor-Agent Actor Critic for Spectrum Allocation}
The spectrum allocation problem formulated in Section \ref{Problem Formulation} can be solved by applying the Q-learning \cite{zia2019distributed}, deep Q Network \cite{ye2019deep} and Actor-Critic (AC) \cite{yang2019intelligent} methods. However, all of the above methods model the reinforcement learning as a MDP. In order to overcome the inherent non-stationary of the multi-agent environment, a Neighbor-Agent Actor-Critic (NAAC) framework is adopted to optimize the policy by modeling multi-agent environment as Markov game and considering action policies of other agents so as to successfully learn policies that require complex multi-agent coordination. In addition, NAAC can make full use of the cooperation between users to further improve the overall performance of the system.

NAAC framework for spectrum allocation in D2D underlay communications is shown in Fig. \ref{NAAC}, which is a framework of centralized training with decentralized execution. A D2D pair $ D_{i} $ is supported by an autonomous agent $ i $. NAAC framework is a extension of AC \cite{lillicrap2015continuous} where each agent is divided into two parts: critic and actor. The actor selects the action according to the observed state, and critic is augmented with extra states and actions information of other neighbor agents to evaluates the quality of the action. We allow the policies to use extra information to ease training, so long as this information is not used at execution time. The centralized training process is done at the BS. In the distributed execution process, a D2D pair $ D_{i} $ (agent $ i $) downloads the trained weight of the actor from the BS and loads it into its own actor $ i $. The actor $ i $ selects action (RB) $ a^{t}_{i} $ based on the state $ s^{t}_{i} $ observed by the agent $ i $ from the environment. When the agent $ i $ takes the action $ a^{t}_{i} $, the environment returns a reward $ r^{t}_{i} $. When the communication is in good condition, the D2D pair $ D_{i} $ can upload the historical information including $ (s^{t}_{i}, a^{t}_{i}, r^{t}_{i}) $ collected at the time of execution to the BS for subsequent training.

A primary motivation behind NAAC is that, if we know the actions taken by all agents, the environment is stationary even as the policies change \cite{lowe2017multi}, since 
\begin{equation}
\begin{aligned}
&p(s_{i}'|s_{i},a_{1},...,a_{N},\pi_{1}, ...,\pi_{N}) 
=p(s_{i}'|s_{i},a_{1},...,a_{N}) \\
&=p(s_{i}'|s_{i},a_{1},...,a_{N},\pi_{1}', ...,\pi_{N}')
\end{aligned}
\end{equation}
for any $ \pi_{i} \neq \pi_{i}' $. The constant transition probability satisfies the Markov assumption of reinforcement learning convergence. In a wireless communication environment, inter-user interference is mainly related to neighbor users. Since when the user's transmit power is constant, the main factor affecting the inter-user interference strength is the large-scale fading, which is mainly related to the distance between users. Therefore, it's no need all users' information to ensure the stability of the environment, just the information of the neighbor users is enough. For a D2D pair $ D_{i} $, we denote a set of $ \lambda $ D2D pairs closest to $ D_{i} $ plus $ D_{i} $ itself as $ \mathcal{N}_{i}^{nb} $. We can use the information of neighbor agents instead of all agents to ensure the stability of the multi-agent environment, since
\begin{equation}
p(s_{i}'|s_{i},a_{1},...,a_{N})
\approx p(s_{i}'|s_{i},\mathbf{a}_{i}^{nb}),
\end{equation}
where $ \mathbf{a}_{i}^{nb} = \{a_{j}, j \in \mathcal{N}_{i}^{nb} \} $ contains the actions of the neighbors of agent $ i $.

The goal in reinforcement learning is to learn a policy $ \pi $ which maximizes the expected return from the start distribution $ J = \mathbb{E}[R^{0}] = \mathbb{E}[\sum _{n=0}^{\infty}\gamma^{n}r^{n}] $. In order to simplify the representation, the state $ s^{t} $, action $ a^{t} $ and return $ R^{t} $ at the current moment are simply denoted as $ s $, $ a $ and $ R $, $ s^{t+1} $ and $ a^{t+1} $ at the next moment are simply denoted as $ s' $ and $ a' $. The action-value function $ Q(s, a) $ corresponds to critic is used in many reinforcement learning algorithms. In a single agent environment, according to AC \cite{lillicrap2015continuous}, consider function approximators parameterized by $ \theta^{Q} $, the critic $ Q(s, a) $ can be optimized by minimizing the loss:
\begin{equation}
Loss(\theta^{Q}) =
\mathbb{E}_{s,a,r,s'}
[(Q(s, a|\theta^{Q}) - y)^{2}],
\end{equation}
where $ y = r + \gamma Q(s', \mu(s')|\theta^{Q}) $, $ \mu $ : $ \mathcal{S} \leftarrow \mathcal{A} $ is a deterministic policy of actor.

Based on Deterministic Policy Gradient \cite{silver2014deterministic}, a parameterized actor function $ \mu(s|\theta^{\mu}) $ can be used to specify the current policy by deterministically mapping states to a specific action. The actor is updated by following the applying the chain rule to the expected return from the start distribution $ J $:
\begin{equation}
\begin{aligned}
\nabla_{\theta^{\mu}} J
\approx 
\mathbb{E}_{s \sim \mathcal{D}}
[\nabla_{a}
Q(s,a|\theta^{Q})|_{a = \mu(s)}]
\nabla_{\theta^{\mu}}\mu(s|\theta^{\mu}),
\end{aligned}
\end{equation}

NAAC extends the AC into multi-agent environment. Consider a Markov game with $ N $ agents, and let $ \bm{\pi} = \{\pi_{1}, ...,\pi_{N}\} $ be the set of all agent policies. We use the states and actions of the neighbor agents of agent $ i $ to evaluate the action-value function (critic) of agent $ i $, which can be written as:
\begin{equation}
Q_{i}^{\bm{\pi}}(\mathbf{s}_{i}^{nb}, \mathbf{a}_{i}^{nb}) =
\mathbb{E}_{\mathbf{s}_{i}^{nb'}, r_{i} \sim E}
[r_{i} + 
\gamma\mathbb{E}_{\mathbf{a}_{i}^{nb'} \sim \bm{\pi}}
Q_{i}^{\bm{\pi}}(\mathbf{s}_{i}^{nb'}, \mathbf{a}_{i}^{nb'})],
\end{equation}
where $ \mathbf{s}_{i}^{nb} $ contains the states of the neighbors of agent $ i $, $ \mathbf{s}_{i}^{nb} = \{ s_{j}, j \in \mathcal{N}_{i}^{nb} \} $, $ Q_{i}^{\bm{\pi}}(\mathbf{s}_{i}^{nb}, \mathbf{a}_{i}^{nb}) $ is a centralized action-value function that takes the states and actions of the agent and its neighbor agents as input, and outputs the Q-value for agent $ i $. 

Let the extended idea work with deterministic policies. If we now consider $ N $ deterministic policies (actor) and let $ \bm{\mu} = \{\mu_{1}, ...,\mu_{N}\} $ be the set of all policies, and the function approximator of the centralized action-value function $ Q_{i} $ of agent $ i $ parameterized by $ \theta_{i}^{Q} $, which we optimize by minimizing the loss:
\begin{equation}
Loss(\theta_{i}^{Q}) =
\mathbb{E}_{\mathbf{s}_{i}^{nb},\mathbf{a}_{i}^{nb},r_{i},\mathbf{s}_{i}^{nb'}}
[(Q_{i}(\mathbf{s}_{i}^{nb}, \mathbf{a}_{i}^{nb}|\theta_{i}^{Q}) - y_{i})^{2}],
\label{NAAC_loss}
\end{equation}
where 
\begin{equation}
y_{i} = r_{i} + \gamma
Q_{i}(\mathbf{s}_{i}^{nb'}, \bm{\mu}(s')|\theta_{i}^{Q})
\label{NAAC_y}.
\end{equation}

Consider the deterministic policy $ \mu_{i} $ of agent $ i $ parameterized by $ \theta_{i}^{\mu} $. We can write the gradient of the expected return for agent $ i $, $ J_{i} = \mathbb{E}[R_{i}] $ as:
\begin{equation}
\begin{aligned}
& \nabla_{\theta_{i}^{\mu}} J_{i}
\approx \mathbb{E}_{\mathbf{s}_{i}^{nb}, \mathbf{a}_{i}^{nb} \sim \mathcal{D}}
[\nabla_{\theta_{i}^{\mu}}
Q_{i}(\mathbf{s}_{i}^{nb}, \mathbf{a}_{i}^{nb}|\theta_{i}^{Q})|
_{a_{i} = \mu(s_{i}|\theta_{i}^{\mu})}] \\
& = \mathbb{E}_{\mathbf{s}_{i}^{nb}, \mathbf{a}_{i}^{nb} \sim \mathcal{D}}
[\nabla_{a_{i}}
Q_{i}(\mathbf{s}_{i}^{nb}, \mathbf{a}_{i}^{nb}|\theta_{i}^{Q})|_{a_{i} = \mu_{i}(s_{i})}]
\nabla_{\theta_{i}^{\mu}}\mu_{i}(s_{i}|\theta_{i}^{\mu}).
\end{aligned}
\label{NAAC_gradient}
\end{equation}
Here $ \mathcal{D} $ is the replay buffer contains the tuples $ (\mathbf{s}, \mathbf{a}, \mathbf{r}, \mathbf{s}') $, recording experiences of all agents. The replay buffer is a finite sized cache. At each timestep the actor and critic are updated by sampling a minibatch uniformly from the buffer.

The mapping between the state and action space of the actor and the action-value function of the critic need to be approximated by function approximators. Q-learning can't work well when the state-action space is very large, where many states may be rarely visited, thus the corresponding Q-values are seldom updated, leading to a much longer time to converge \cite{ye2019deep}. Therefore, deep learning is introduced into NAAC, and deep neural networks (DNNs) are used to approximate the mapping in high-dimensional space.

In NAAC, we denote the set of actor networks and critic networks of all agents as $ \bm{\mu} = \{\mu_{1}, ...,\mu_{N}\} $ and $ \mathbf{Q} = \{Q_{1}, ...,Q_{N}\} $ with the weights $ \bm{\theta}^{\bm{\mu}} = \{\theta_{1}^{\mu}, ...,\theta_{N}^{\mu}\} $ and $ \bm{\theta}^{\mathbf{Q}} = \{\theta_{1}^{Q}, ...,\theta_{N}^{Q}\} $, respectively. The input of the actor network is the state observed by the agent, and the output is the selected action. The hidden layers in the actor network are all fully connected layers. For the critic network, first enter $ \mathbf{s}_{i}^{nb} $, after a fully connected layer, then enter $ \mathbf{a}_{i}^{nb} $, then go through several fully connected layers, and finally output $ Q_{i}^{\bm{\pi}}(\mathbf{s}_{i}^{nb}, \mathbf{a}_{i}^{nb}) $ . 

\subsection{Training Algorithm}
Since BS has more computing power than the mobile device, the centralized training process is done at the BS, As shown in the green part of Figure \ref{NAAC}. The training algorithm is shown in Algorithm 1. The NAAC framework uses historical information to train the DNNs of actor and critic, and returns the weights of actor network $ \bm{\theta}^{\bm{\mu}} $.

\begin{algorithm}\small
	\caption{Training Algorithm}  
	\label{Training}
	\textbf{Input:} Actor network structure, critic network structure. \\
	\textbf{Output:} The weights of actor network $ \bm{\theta}^{\bm{\mu}} $. \\
	\textbf{Training:}
	\begin{algorithmic}[1]
		\STATE {Random initialize actor network $ \bm{\mu} $ and critic network $ \mathbf{Q} $ with the weights $ \bm{\theta}^{\bm{\mu}} $ and $ \bm{\theta}^{\mathbf{Q}} $.}
		\STATE {All agents (D2D pairs) receive initial observation states $ \mathbf{s}^{0} = \{s_{1}^{0}, ..., s_{N}^{0}\} $} 
		\FOR{each time slot $ t=0,1,...T $.}
		\STATE {All agents select actions 
			$ \mathbf{a}^{t} = \{a_{i}^{t} = \mu_{i}(s_{i}^{t}), i \in \mathcal{N}\} $ according to the current policy.}
		\STATE {All agents take actions $ \mathbf{a}^{t} $, observe rewards $ \mathbf{r}^{t} = \{r_{1}^{t}, ..., r_{N}^{t}\} $ and observe new states $ \mathbf{s}^{t+1} $.}
		\STATE {Save the tuples $ (\mathbf{s}^{t}, \mathbf{a}^{t}, \mathbf{r}^{t}, \mathbf{s}^{t+1}) $ in $ \mathcal{D} $.} 
		\STATE {Sample a random mini-batch of tuples $ (\mathbf{s}, \mathbf{a}, \mathbf{r}, \mathbf{s}') $ from $ \mathcal{D} $.}
		\STATE {Set $ y_{i} = r_{i} + \gamma
			Q_{i}(\mathbf{s}_{i}^{nb'}, \bm{\mu}(s'|\bm{\theta}^{\bm{\mu}})|\theta_{i}^{Q}) $.}
		\STATE {Update critic by minimizing the loss in equation (\ref{NAAC_loss}).}
		\STATE {Update the actor policy according to equation (\ref{NAAC_gradient}).}
		\ENDFOR
	\end{algorithmic} 
\end{algorithm}

\section{Performance Evaluation}
In this section, we present the simulation results of the NAAC in comparison to four distributed approaches: 1. The most classic reinforcement learning method Q-learning \cite{zia2019distributed}; 2. A reinforcement learning method with better convergence performance, Actor-Critic (denoted as AC) \cite{yang2019intelligent}; 3. The most classic deep reinforcement learning method Deep Q Network (denoted as DQN) \cite{ye2019deep}; 4. A game theory approach, Uncoupled Stochastic Learning Algorithm developed in \cite{dominic2018distributed} (denoted as SLA). Since we assume that each D2D pair can only obtain its own CSI and there is no information interaction between D2D users, centralized approaches with global information do not participate in performance comparisons.

For the simulations, we consider a single cell scenario with a radius of 500m. The CUEs and D2D pairs are distributed randomly in a cell, where the communication distance of each D2D pair cannot exceed a given maximum distance 30m. The detail parameters can be found in Table \ref{parameters}.

\subsection{Simulations Results}

\begin{table}[!t]
	\caption{SIMULATION PARAMETERS}
	\begin{center}
		\begin{tabular}{|c|c|}
			\hline
			\textbf{Parameter}& \textbf{Value} \\
			\hline
			RB bandwidth & 180 KHz\\
			\hline
			Number of CUEs & 10\\
			\hline
			Number of RBs & 10\\
			\hline
			BS transmission power ($ P^{b} $) & 46 dBm\\
			\hline
			D2D transmission power ($ P^{d} $) & 13 dBm\\
			\hline
			Cellular link pathloss & $ 128.1+37.6 \log _{10} (d[km]) $\\
			\hline
			D2D link path loss exponent & 4\\
			\hline
			UE thermal noise density & -174 dBm/Hz\\
			\hline
			CUE target SINR threshold ($ \xi^{c}_{min} $) & 0 dB\\
			\hline
			Negative reward ($ r_{neg} $) & -1 \\
			\hline
		\end{tabular}
		\label{parameters}
	\end{center}
\end{table}

\begin{figure}[!t]
	\centering
	\includegraphics[width=3.5in]{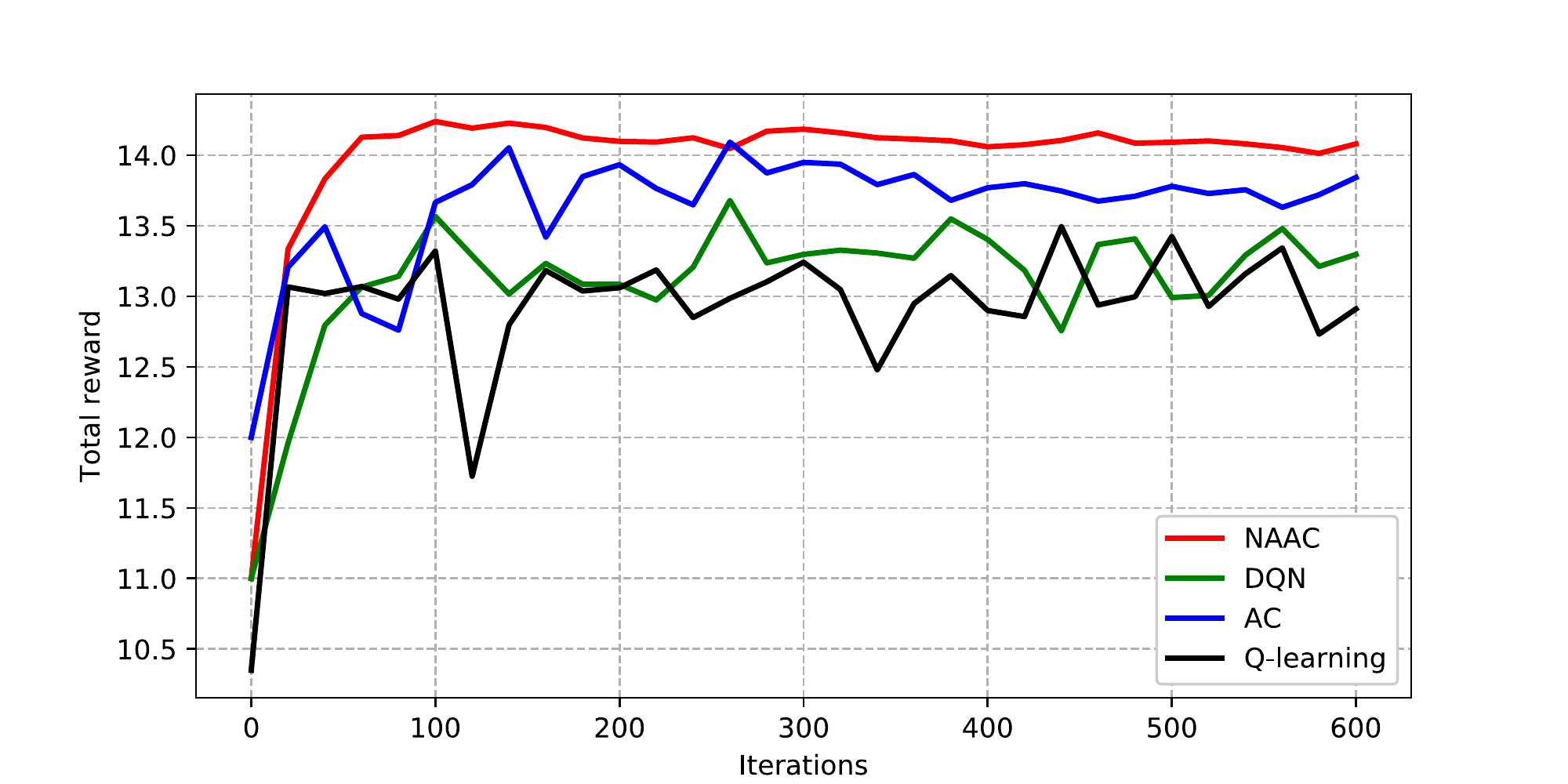}
	\caption{Comparison of total reward performance during training process.}
	\label{reward}
\end{figure}

In order to compare the convergence of these algorithms. Fig. \ref{reward} shows the training process of the four approaches in terms of the total reward performance when the number of D2D pairs is 10. Total reward is the sum of the rewards obtained by all agents. Since SLA is an online learning algorithm, it does not participate in the discussion of this indicator. We can see that NAAC has achieved the greatest total reward, while the convergence is most stable. The total reward and convergence of Q-learning are the worst, since Q-learning can't work well when the state-action space is vary large. DQN solves the mapping problem of high-dimensional space by using DNN, improves in both total reward performance and convergence than Q-learning. For the AC, its performance is better than Q-learning and DQN, since AC optimizes the policy by combining the process of the policy learning and value learning with good convergence properties. However, none of the above algorithms considers the impact of multi-agent environment on stability of training process and the cooperation between multiple agents (D2D pairs) on system performance. The NAAC introduces the state and action information of neighbor D2D pairs to assist the training process, greatly improving the stability of the training process, and achieving a higher total reward.

\begin{figure}[!t]
	\centering
	\includegraphics[width=3.5in]{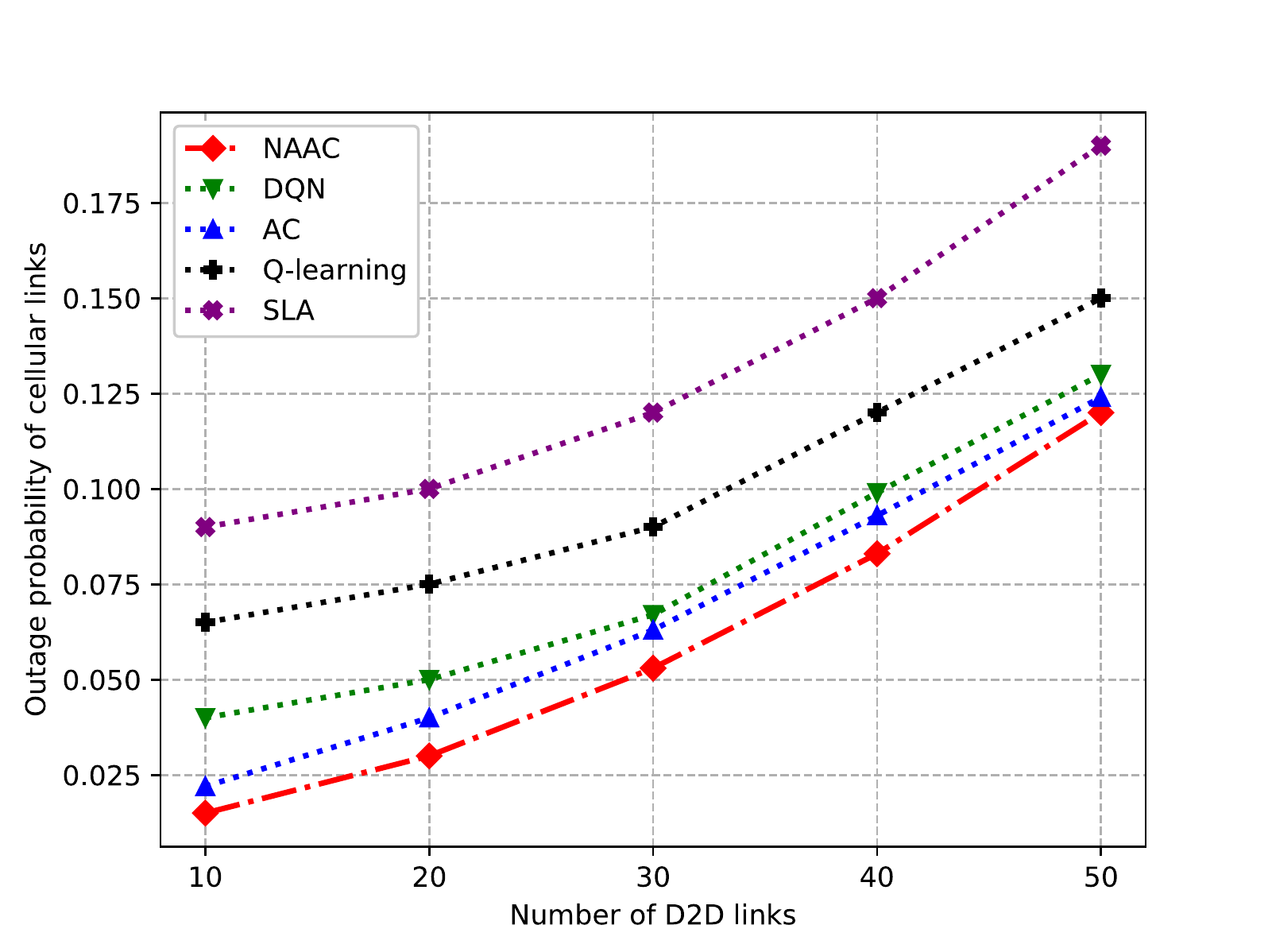}
	\caption{Outage probability of cellular links versus the number of D2D links.}
	\label{cue_op}
\end{figure}

The outage probability can reflect the reliability of the communication links. In Fig. \ref{cue_op}, we show the outage probability of cellular links as a function of the number of D2D pairs $ N $, and NAAC works with the number of neighbor users $ \lambda=3 $. The outage probability of cellular links increases as the number of D2D links grows, since there are more D2D pairs sharing the spectrum with CUEs, which causes the CUEs to suffer more severe cross-layer interference. It's shown that NAAC achieved the best performance, since the other four methods update their own policies independently during the training, NAAC can learn the policy of cooperation among users by introducing the states and actions of neighbor D2D pairs to assist the training. Therefore, the policies between different D2D pairs can be coordinated with each other to prevent multiple D2D pairs from simultaneously selecting the same RB to cause severe cumulative interference to the CUE.

\begin{figure}[!t]
	\centering
	\includegraphics[width=3.5in]{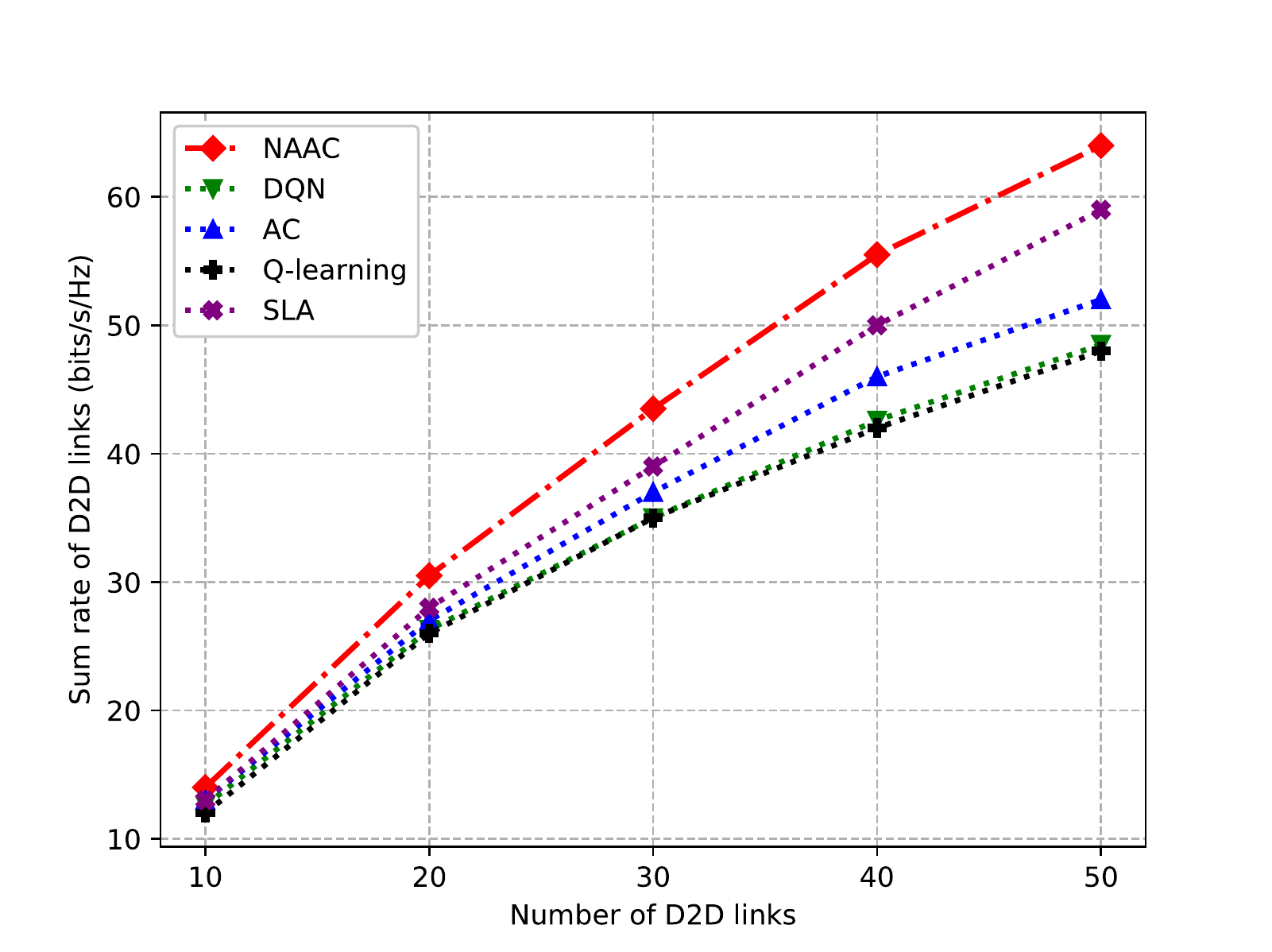}
	\caption{Sum rate of D2D links versus the number of D2D links.}
	\label{d2d_sum_rate}
\end{figure}

Fig. \ref{d2d_sum_rate} shows the D2D sum rate as a function of $ N $, and NAAC works with $ \lambda=3 $. The the D2D sum rate increases as the number of D2D links grows, since more D2D pairs are allocated to RBs. It can be seen that NAAC is significantly better than the other four algorithms, and as the number of D2D links increases, the advantages become larger. Since the other four algorithms can only achieve individual optimization, the effect of global optimization cannot be guaranteed, but NAAC of centralized training with decentralized execution can cooperatively optimize the sum rate of D2D links.

\subsection{Discussion}
The proposed framework combines the strengths of centralized and distributed schemes. Compared with the centralized method, our method is executed without global information, which significantly reduces the signaling overhead and alleviates the computational pressure of the BS. Therefore, our method can scale well to larger networks. Compared with distributed method, our method uses the historical information of neighbor users to learn the policies of mutual cooperation, avoiding frequent real-time information exchange between users, more suitable for user-intensive communication scenario. In addition, our method can transfer complex training processes to the cloud (BS), significantly reduces the complexity of algorithm execution.

\section{Conclusion}
This paper has studied the resource management in D2D underlay communications, and formulated the intelligent spectrum allocation problem as a decentralized multi-agent deep reinforcement learning problem. For full cooperation between users, the NAAC framework of centralized training with distributed execution is adopted, which requires no signaling interaction. The simulation results show that the proposed approach can effectively reduce the outage probability of cellular links, improve the sum rate of D2D links and have good convergence.

\bibliographystyle{ieeetr}
\footnotesize
\bibliography{ref}

\begin{thebibliography}{10}

\bibitem{kai2019resource}
Y.~Kai, J.~Wang, H.~Zhu, and J.~Wang, ``Resource allocation and performance
  analysis of cellular-assisted ofdma device-to-device communications,'' {\em
  IEEE Transactions on Wireless Communications}, vol.~18, no.~1, pp.~416--431,
  2019.

\bibitem{feng2013device}
D.~Feng, L.~Lu, Y.~Yuan-Wu, G.~Y. Li, G.~Feng, and S.~Li, ``Device-to-device
  communications underlaying cellular networks,'' {\em IEEE Transactions on
  Communications}, vol.~61, no.~8, pp.~3541--3551, 2013.

\bibitem{wu2018high}
D.~Wu and N.~Ansari, ``High capacity spectrum allocation for multiple d2d users
  reusing downlink spectrum in lte,'' in {\em 2018 IEEE International
  Conference on Communications (ICC)}, pp.~1--6, IEEE, 2018.

\bibitem{kose2018resource}
A.~K{\"o}se and B.~{\"O}zbek, ``Resource allocation for underlaying
  device-to-device communications using maximal independent sets and knapsack
  algorithm,'' in {\em 2018 IEEE 29th Annual International Symposium on
  Personal, Indoor and Mobile Radio Communications (PIMRC)}, pp.~1--5, IEEE,
  2018.

\bibitem{kuang2019energy}
Z.~Kuang, G.~Liu, G.~Li, and X.~Deng, ``Energy efficient resource allocation
  algorithm in energy harvesting-based d2d heterogeneous networks,'' {\em IEEE
  Internet of Things Journal}, vol.~6, no.~1, pp.~557--567, 2019.

\bibitem{zaki2017distributed}
F.~Zaki, S.~Kishk, and N.~Almofari, ``Distributed resource allocation for {D2D}
  communication networks using auction,'' in {\em 2017 34th National Radio
  Science Conference (NRSC)}, pp.~284--293, IEEE, 2017.

\bibitem{jiang2017machine}
C.~Jiang, H.~Zhang, Y.~Ren, Z.~Han, K.-C. Chen, and L.~Hanzo, ``Machine
  learning paradigms for next-generation wireless networks,'' {\em IEEE
  Wireless Communications}, vol.~24, no.~2, pp.~98--105, 2017.

\bibitem{sutton1998introduction}
R.~S. Sutton, A.~G. Barto, {\em et~al.}, {\em Introduction to reinforcement
  learning}, vol.~135.
\newblock MIT press Cambridge, 1998.

\bibitem{zia2019distributed}
K.~Zia, N.~Javed, M.~N. Sial, S.~Ahmed, A.~A. Pirzada, and F.~Pervez, ``A
  distributed multi-agent {RL}-based autonomous spectrum allocation scheme in
  {D2D} enabled multi-tier {HetNets},'' {\em IEEE Access}, vol.~7,
  pp.~6733--6745, 2019.

\bibitem{yang2019intelligent}
H.~Yang, X.~Xie, and M.~Kadoch, ``Intelligent resource management based on
  efficient transfer actor-critic reinforcement learning for {IoV}
  communication networks,'' {\em IEEE Transactions on Vehicular Technology},
  2019.

\bibitem{ye2019deep}
H.~Ye, Y.~G. Li, and B.-H.~F. Juang, ``Deep reinforcement learning for resource
  allocation in {V2V} communications,'' {\em IEEE Transactions on Vehicular
  Technology}, 2019.

\bibitem{plaisted1976some}
D.~A. Plaisted, ``Some polynomial and integer divisibility problems are
  {NP-hard},'' in {\em 17th Annual Symposium on Foundations of Computer Science
  (sfcs 1976)}, pp.~264--267, IEEE, 1976.

\bibitem{lillicrap2015continuous}
T.~P. Lillicrap, J.~J. Hunt, A.~Pritzel, N.~Heess, T.~Erez, Y.~Tassa,
  D.~Silver, and D.~Wierstra, ``Continuous control with deep reinforcement
  learning,'' {\em arXiv preprint arXiv:1509.02971}, 2015.

\bibitem{lowe2017multi}
R.~Lowe, Y.~Wu, A.~Tamar, J.~Harb, O.~P. Abbeel, and I.~Mordatch, ``Multi-agent
  actor-critic for mixed cooperative-competitive environments,'' in {\em
  Advances in Neural Information Processing Systems}, pp.~6379--6390, 2017.

\bibitem{silver2014deterministic}
D.~Silver, G.~Lever, N.~Heess, T.~Degris, D.~Wierstra, and M.~Riedmiller,
  ``Deterministic policy gradient algorithms,'' in {\em ICML}, 2014.

\bibitem{dominic2018distributed}
S.~Dominic and L.~Jacob, ``Distributed resource allocation for {D2D}
  communications underlaying cellular networks in time-varying environment,''
  {\em IEEE Communications Letters}, vol.~22, no.~2, pp.~388--391, 2018.

\end{thebibliography}
\end{document}